\documentclass[twocolumn,aps,prd,showpacs,preprintnumbers,nofootinbib]{revtex4}
\usepackage[dvips]{graphicx}
\usepackage{bm,latexsym,amsmath,amssymb,amsfonts}
\begin{document}

\title{No-dipole-hair theorem for higher-dimensional static black holes}

\author{Roberto Emparan}
\affiliation{Instituci\'o Catalana de Recerca i Estudis
Avan\c cats (ICREA), Passeig Llu\'{\i}s Companys 23, E-08010 Barcelona, Spain}
\affiliation{Departament de F{\'\i}sica Fonamental and Institut de
Ci\`encies del Cosmos, Universitat de
Barcelona, Mart\'{\i} i Franqu\`es 1, E-08028 Barcelona, Spain}
\author{Seiju Ohashi and Tetsuya Shiromizu}
\affiliation{Department of Physics, Kyoto University, Kyoto 606-8502, Japan}
\begin{abstract}
We prove that static black holes in $n$-dimensional 
asymptotically flat spacetime cannot support nontrivial electric $p$-form
field strengths when $(n+1)/2\leq p\leq n-1$. This implies, in particular, that static
black holes cannot possess dipole hair under these fields.
\end{abstract}
\maketitle

\section{Introduction}

It has been known for several years that the properties of
higher-dimensional black holes can differ significantly from the rigidly
constrained character of four-dimensional black holes
\cite{Emparan:2008eg}.
References.~\cite{Emparan:2003sy,Emparan:2009cs} trace the
origin of the new physics back to the possibility of having two
parametrically different length scales along the horizon, which allows
higher-dimensional black holes to exhibit black brane-like behavior that
cannot occur in four dimensions. Typically, one of the scales is
associated to the black hole mass and the other to its angular momentum,
which in five or more dimensions can be arbitrarily large for a given
mass. As the two scales begin to separate, new phenomena set in, such as
black hole nonuniqueness and horizon instabilities. 

This observation suggests that the distinctively higher-dimensional
features of black holes arise only at sufficiently large rotation. In
particular, it leads us to expect that the properties of static black
holes should be qualitatively very similar to those of four-dimensional
black holes. There is already good evidence for this. Prompted by the
discovery of black rings and the nonuniqueness of stationary black
holes that they entail \cite{Emparan:2001wn}, Ref.~\cite{GIS1} showed
that asymptotically flat, static vacuum black holes are instead unique:
the only solution is the Schwarzschild-Tangherlini spacetime. Afterward
it was also shown that this solution is dynamically stable to linearized
perturbations \cite{Ishibashi:2003ap}. While we expect that uniqueness
(for solutions with connected horizons) and stability are valid in a
finite range of values of the angular momenta, the precise upper limits
are still unknown in general (see \cite{Dias:2009iu,Shibata:2009ad} for
some recent progress in this direction).

It is clearly of interest to study how these results are extended when
gauge fields are present. A charge on the black hole introduces an
independent length scale, namely the charge radius. A separation of
scales occurs as extremality is approached, but this occurs in
directions transverse to the horizon instead of parallel to it, and is
in fact an effect well-known in four dimensions too. So, again, the
onset of qualitatively new higher-dimensional features seems to require
a minimum amount of rotation. In this direction, Ref.~\cite{GIS2} has
proven the uniqueness of the $n$-dimensional static black holes
electrically charged under a two-form field strength, and
Ref.~\cite{Kodama:2003kk} has studied their stability.

A more distinctive property of gauge fields in higher dimensions is the
possibility that a black hole couples electrically to $p$-form field
strengths $H_{(p)}$ with $p>2$. An asymptotically flat black hole in $n$
dimensions cannot have a conserved monopolar electric charge under this
field. One might expect that the integral of $\ast H_{(p)}$ over a
sphere $S^{n-p}$ near asymptotically flat infinity gives a conserved
charge. However, this is not the case when $p>2$, since the $S^{n-p}$
can be shrunk to a point in the $S^{n-2}$ in the asymptotically flat
region and the integral vanishes. Nevertheless, the black hole can be the
source of an electric dipole of this field. Indeed,
Ref.~\cite{Emparan:2004wy} presented rotating black ring solutions with
dipoles of a three-form field strength in five dimensions. Since the dipole
is not a conserved charge, it is hair for the black hole. The generic
existence of rotating black holes with dipoles of fields $H_{(p)}$ with
$p\geq 3$ in dimensions $n\geq p+2$ is argued in \cite{EHNOip}. 

Could a static black hole sport such dipole hairs? Intuitively, the
dipole field can be regarded as sourced by a $(p-2)$-brane-like object
extended along a compact $(p-2)$-cycle. This
``brane'' exerts a tension that, if the cycle is contractible, must be
balanced by centrifugal rotation (this is indeed explicitly observed in
the dipole rings of \cite{Emparan:2004wy}). So this heuristic reasoning
indicates that we should not expect a black hole to be able to support a
dipole until it carries a sufficiently large angular momentum. 

In this article we prove the impossibility of dipole hair for static
black holes. The proof follows the one employed in the uniqueness
theorem of higher-dimensional static black holes
\cite{GIS1,Hwang,GIS2,Rogatko1}. This type of proof was first developed
by Bunting and Masood-ul-Alam in four dimensions \cite{BM}. Its
extension to higher dimensions is quite nontrivial, since \cite{BM}
used properties specific to four dimensions, but the approach was
extended in \cite{GIS1,Hwang} in a manner that avoids the use of such
properties. 

Together with the gauge dipole, we will also consider the inclusion
of scalar fields and scalar hair. Bekenstein proved that a static black
hole can not have scalar hair in four dimensions \cite{BS}. This no-hair
theorem is easily extended to higher dimensions since the dimensionality
does not enter into the proof. However, this type of proof cannot be applied to systems where 
the scalar field couples to higher form fields. 

The rest of this paper is organized as follows. In Sec.~II, we prove the
no-dipole-hair theorem in two steps: first we show that a static black
hole cannot support nontrivial $p$-form fields when $(n+1)/2\leq p\leq
n-1$, and then we prove the uniqueness of the
Schwarzschild-Tangherlini solution. In Sec.~III, we discuss the outlook
of this work, and in particular the restriction on
the values of $p$ to which the theorem applies. 

\section{No-dipole-hair theorem}

We consider $n$-dimensional asymptotically flat solutions of
theories described by the class of Lagrangians
%============<Equation>=============%
%
\begin{equation}
{\cal L}=R-\frac{1}{2}(\partial \phi)^2-\frac{1}{p!}e^{-\alpha \phi}H_{(p)}^2,\label{Lag}
\end{equation} 
%
%===================================%
where $R$ is the $n$-dimensional Ricci scalar, $\phi$ is a dilaton with
coupling $\alpha$, and $H_{(p)}$ 
is the field strength of a $(p-1)$-form field potential $B_{(p-1)}$,
%============<Equation>=============%
%
\begin{equation}
H_{(p)}=dB_{(p-1)}.
\end{equation} 
%
%===================================%
Since we are interested in asymptotically flat spacetimes, we take $p\leq n-1$. 
A form field with $p=n$ does not have any dynamical degree 
of freedom and behaves like a cosmological constant, which would prevent
asymptotic flatness. 

We only consider electric fields of $H_{(p)}$.
Note that via electric-magnetic duality we can always trade a magnetic
charge or dipole under $H_{(p)}$ for an electric one under $H_{(n-p)}$
\footnote{Reference \cite{Rogatko2} purports to study black holes with both
electric and magnetic monopole charges under $H_{(n-2)}$, but if $n>4$
this is impossible for the reasons given above.}. However, we do not
consider the possibility of simultaneous presence of dipoles and
monopole charges of electric and magnetic type, \textit{e.g.}, in
$n=p+2$ one can have solutions with both magnetic monopole charge and
electric dipole of $H_{(p)}$. Sometimes these involve an additional
Chern-Simons term in the action, which however is inconsequential
for our analysis involving only electric fields. The uniqueness of
$U(1)^2$-symmetric black holes in some such theories in five dimensions
has been discussed in \cite{othercharge}.

The equations of motion for the theories (\ref{Lag}) are 
%============<Equation>=============%
%
\begin{equation}
\nabla^2 \phi=-\frac{\alpha}{p!}e^{-\alpha\phi}H_{(p)}^2
\end{equation}
%
%===================================%  
%============<Equation>=============%
%
\begin{equation}
\nabla_M (e^{-\alpha\phi}H^{MN_1 \cdots N_{p-1}})=0
\end{equation}
%
%===================================%  
and
%============<Equation>=============%
%
\begin{eqnarray}
R_{MN} & = & \frac{1}{2}\nabla_M \phi \nabla_N \phi+\frac{1}{p!}
e^{-\alpha\phi}\Bigl(pH_M^{~~I_1 \cdots I_{p-1}}H_{NI_1 \cdots I_{p-1}} \nonumber \\
& & -\frac{p-1}{n-2}g_{MN}H_{(p)}^2 \Bigr),
\end{eqnarray}
%
%===================================%  
where $\nabla_M$ is the covariant derivative with respect to $g_{MN}$,
and $M,N=0,\dots n-1$. 

The metric of a static spacetime can be written as 
%============<Equation>=============%
%
\begin{equation}
ds^2=g_{MN}dx^M dx^N=-V^2(x^i)dt^2+g_{ij}(x^k)dx^i dx^j,
\end{equation}
%
%===================================%  
where $x^i$ are spatial coordinates on $x^0=t=$const. surfaces $\Sigma$. In these coordinates, the 
event horizon is located at $V=0$, {\it i.e.}, the Killing horizon. 
The static ansatz for the $(p-1)$-form potential is of the form
%============<Equation>=============%
%
\begin{eqnarray}
B_{(p-1)}=\varphi_{i_1 \cdots i_{p-2}}(x^k) dt \wedge dx^{i_1} \wedge 
\cdots  \wedge  dx^{i_{p-2}}.
\end{eqnarray}
%
%===================================%
Then the only nontrivial component of the field strength is $H_{0 i_1 \cdots 1_{p-1}}$. 
The metric components and the potential do not depend on $t$. 

We shall prove the following theorem:\\

{\it No-dipole-hair theorem: The only static, asymptotically flat black
hole solution for the theories (\ref{Lag}) with electric $p$-form
field strength, with $(n+1)/2 \leq p\leq n-1$, is the
Schwarzschild-Tangherlini solution. }\\

From the Einstein equation we have 
%============<Equation>=============%
%
\begin{eqnarray}
R_{ij} & = & {}^{(n-1)}R_{ij}-\frac{1}{V}D_i D_j V \nonumber \\
& = & \frac{1}{2}D_i \phi D_j \phi \nonumber \\
& & +
\frac{1}{(p-2)!}e^{-\alpha \phi}
\Bigl(H_i^{~~0k_1 \cdots k_{p-2}}H_{j0k_1 \cdots k_{p-2}} \nonumber \\
& & -\frac{1}{n-2}g_{ij}H_{0k_1 \cdots k_{p-1}}H^{0k_1 \cdots k_{p-1}} \Bigr)
\end{eqnarray} 
%
%===================================% 
and
%============<Equation>=============%
%
\begin{eqnarray}
R_{00} & = & VD^2V\nonumber \\
& =  &\frac{n-p-1}{(n-2)(p-1)!}e^{-\alpha\phi}
H_0^{~~i_1 \cdots i_{p-1}}H_{0i_1 \cdots i_{p-1}},
\end{eqnarray} 
%
%===================================% 
where $D_i$ is the covariant derivative with respect to $g_{ij}$. 
From these we derive 
%============<Equation>=============%
%
\begin{equation}
{}^{(n-1)}R=\frac{e^{-\alpha\phi}}{(p-1)!V^2}H_0^{~~i_1 \cdots i_{p-1}}H_{0i_1 \cdots i_{p-1}} 
+\frac{1}{2}(D\phi)^2 \label{n-1ricci}
\end{equation} 
%
%===================================% 
and
%============<Equation>=============%
%
\begin{eqnarray}
D^2V=\frac{n-p-1}{(n-2)(p-1)!} \frac{e^{-\alpha\phi}}{V}H_0^{~~i_1 \cdots i_{p-1}}
H_{0i_1 \cdots i_{p-1}}.
\end{eqnarray} 
%
%===================================% 
From the equations for the form field we obtain 
%============<Equation>=============%
%
\begin{equation}
D_i (e^{-\alpha \phi}H^i_{~~j_1 \cdots j_{p-2}0})=\frac{D_iV}{V}e^{-\alpha \phi}H^i{}_{j_1 \cdots j_{p-2}0} .
\end{equation} 
%
%===================================% 

The asymptotic behavior of $V, g_{ij},$ and $H_{(p)}$ is 
%============<Equation>=============%
%
\begin{eqnarray}
& & V=1-\frac{m}{r^{n-3}}+O(1/r^{n-2})\\
& & g_{ij}=\delta_{ij}\Bigl(1+ \frac{2}{n-3}\frac{m}{r^{n-3}} \Bigr)+O(1/r^{n-2}) \\
& & H_{0i_1 \cdots i_{p-1}}=O(1/r^{n-p+1}). 
\end{eqnarray} 
%
%===================================% 
Observe that the falloff of $H_{(p)}$ is the appropriate one for a
dipole field, or higher multipole components. In our proof this decay
rate could be relaxed to that of a monopole field, $O(1/r^{n-p})$.
However, as we explained in the introduction, when $p>2$ electric
monopole charges are incompatible with asymptotic flatness. 

We also assume regularity on the event horizon. To this effect, 
we compute the curvature invariant 
%============<Equation>=============%
%
\begin{eqnarray}
& & R_{MNKL}R^{MNKL} \nonumber \\
& & ~~= {}^{(n-1)}R_{ijkl}{}^{(n-1)}R^{ijkl}+
4{}^{(n-1)}R_{0i0j}{}^{(n-1)}R^{0i0j} \nonumber \\
& & ~~=  {}^{(n-1)}R_{ijkl}{}^{(n-1)}R^{ijkl}+4\frac{D_i D_j V D^i D^j V}{V^4}\nonumber \\
& & ~~=  {}^{(n-1)}R_{ijkl}{}^{(n-1)}R^{ijkl} \nonumber \\
& & ~~~~+\frac{4(n-2)}{(n-3)V^2\rho^2}[k_{ab}k^{ab}+k^2+{\cal D}_a \rho {\cal D}^a \rho].
\label{inv}
\end{eqnarray} 
%
%===================================% 
Here we have used the fact that the spatial metric can be 
written as
%============<Equation>=============%
%
\begin{eqnarray}
g_{ij}dx^i dx^j =\rho^2dV^2+h_{ab}dx^adx^b,
\end{eqnarray} 
%
%===================================% 
where $x^a$ is the coordinate on the level surfaces of $V$. 
${\cal D}_a$ is the covariant derivative with respect to $h_{ab}$. 
$k_{ab}$ is the extrinsic curvature of $V=$const. surface and $\rho:=|D^iV D_iV|^{-1/2}$. 
Then, from Eq. (\ref{inv}), one can easily see that 
%============<Equation>=============%
%
\begin{eqnarray}
k_{ab}|_{V=0}={\cal D}_a \rho|_{V=0}=0
\end{eqnarray} 
%
%===================================% 
hold on the event horizon. From the Einstein equation, we can also easily see 
that regularity implies $H_{0i_1 \cdots i_{p-1}}=0$ on the event 
horizon; see Eq.~(\ref{n-1ricci}). 

Let us consider the conformal transformation defined by 
%============<Equation>=============%
%
\begin{eqnarray}
\tilde g_{ij}=\Omega_\pm^2 g_{ij}
\end{eqnarray} 
%
%===================================% 
where
%============<Equation>=============%
%
\begin{equation}
\Omega_\pm=\Biggl( \frac{1\pm V}{2} \Biggr)^{\frac{2}{n-3}}=:\omega_\pm^{\frac{2}{n-3}}.
\label{confOm}
\end{equation} 
%
%===================================% 
This conformal transformation is the same as the one employed in the proof for 
the vacuum case \cite{GIS1, Hwang}. 
Now we have two manifolds, $(\tilde \Sigma^+,\tilde g^+)$ and $(\tilde \Sigma^-,\tilde g^-)$. 
The Ricci scalar of $\tilde \Sigma^\pm$ is 
%============<Equation>=============%
%
\begin{eqnarray}
& & \Omega_\pm^2 {}^{(n-1)}\tilde R_\pm \nonumber \\\
& &=  {}^{(n-1)}R-2(n-2)D^2 \ln \Omega_\pm \nonumber \\
& & ~~~~-(n-3)(n-2) (D \ln \Omega_\pm)^2\nonumber \\
& &=  {}^{(n-1)}R \mp \frac{2(n-2)}{n-3} \omega_{\pm}^{-1}D^2V \nonumber \\
& & = \frac{1}{(p-1)!}\frac{e^{-\alpha\phi}}{V^2}\frac{\lambda_\pm}{\omega_\pm}H_0^{~~i_1 \cdots i_{p-1}}
H_{0i_1 \cdots i_{p-1}}+\frac{1}{2}(D\phi)^2, \nonumber \\
& & 
\label{OmR}
\end{eqnarray} 
%
%===================================% 
where 
%============<Equation>=============%
%
\begin{eqnarray}
\lambda_\pm :=\frac{1\mp\frac{3n-4p-1}{n-3} V}{2}. 
\end{eqnarray}
%
%===================================% 
Since $0\leq V \leq 1$, the $\lambda_\pm$ are positive-definite if 
%============<Equation>=============%
%
\begin{eqnarray}
\frac{n+1}{2} \leq p\leq n-1. \label{p-condition}
\end{eqnarray} 
%
%===================================% 
Under this condition the positivity of 
${}^{(n-1)}\tilde R_\pm$ follows. 
We will use this result later. 

On $\tilde \Sigma^+$ the asymptotic behavior of the metric becomes 
%============<Equation>=============%
%
\begin{equation}
\tilde g_{ij}^+=\Bigl(1+O( 1/r^{n-2}) \Bigr) \delta_{ij}
\end{equation} 
%
%===================================% 
and therefore the ADM mass vanishes there. On $\tilde \Sigma^-$, 
the metric behaves like 
%============<Equation>=============%
%
\begin{eqnarray}
\tilde g_{ij}^-dx^idx^j
& = & \frac{(m/2)^{4/(n-3)}}{r^4}\delta_{ij}dx^i dx^j+O(1/r^5) \nonumber \\
& = & (m/2)^{4/(n-3)} (d\rho^2+\rho^2d\Omega^2_{n-2})+O(\rho^5),\nonumber \\
& & 
\end{eqnarray} 
%
%===================================% 
where we set $\rho:=1/r$. From this, we see that 
infinity on $\Sigma$ corresponds to a point, which we denote as $q$. 

Let us construct a new manifold $(\tilde \Sigma, \tilde g_{ij})
:=(\tilde \Sigma^+,\tilde g^+_{ij}) \cup (\tilde \Sigma^-, \tilde
g^-_{ij}) \cup \lbrace q \rbrace $ by gluing the two manifolds $(\tilde
\Sigma^+,\tilde g^+_{ij})$ and $(\tilde \Sigma^-, \tilde g^-_{ij})$
along the surface $V=0$ and adding the point $q$.\footnote{Note that the
resulting manifold $\tilde \Sigma$ is $C^1$ on the surface $V=0$. This
is as in the vacuum case \cite{GIS1}, since the conformal transformation
is the same, as mentioned above.} The calculations above imply that
$(\tilde \Sigma, \tilde g_{ij})$ has zero mass and non-negative Ricci
scalar. Note also that near the point $q$ (which corresponds to $r \to
\infty$) we have ${}^{(n-1)}\tilde R_- =O(r^{-(n-3)})$, so $\tilde
\Sigma^-$ is regular at $q$ \footnote{Even if the system has monopole
charge, ${}^{(n-1)}\tilde R_- =O(r^{-(n-5)})$. So $\tilde \Sigma^-$ will
be regular whenever $n\geq 5$. Our method of proof does not
depend on which hair the system has. If $H_{0 i_1 \cdots
i_{p-1}}=O(1/r^s)$, ${}^{(n-1)}\tilde R_- =O(r^{n-2p-2s+5})$. Then
regularity requires $s \geq (n+1)/2-(p-2)$.}. Thus $\tilde\Sigma$ is a
Riemannian manifold with non-negative Ricci scalar and zero ADM mass.
Then, by the positive energy theorem \cite{PET}, $\tilde\Sigma$ is flat.
So the metric $\tilde g_{ij}$ is flat and 
%============<Equation>=============%
%
\begin{equation}
H_{0i_1 \cdots i_{p-1}}=0~~{\rm and~~}\phi={\rm const}
\end{equation} 
%
%===================================% 
hold\footnote{More precisely, this argument requires that the surface
$t=$const.\ is a spin-manifold or a manifold of dimension less than eight \cite{PET2}.}. That is, asymptotically flat static black holes in
$n$ dimensions cannot support an electric dipole $p$-form field
strength with $p$ in the range (\ref{p-condition}), nor a nontrivial
scalar field.

Once we have ruled out the possibility of nontrivial $p$-form and
scalar fields, the problem is exactly the same as in vacuum and the results
of \cite{GIS1} imply the uniqueness of the Schwarzschild-Tangherlini
solution. For the sake of completeness, we briefly review this argument.

We have seen that $\tilde \Sigma^+$ must be flat space. In addition, we
can check that the extrinsic curvature of the surface $V=0$ on $\tilde
\Sigma^+$ is proportional to its induced metric with a constant
coefficient. According to Kobayashi and Nomizu \cite{KN}, such a surface
in flat space is spherically symmetric. Next, we define the function $v$
by 
%============<Equation>=============%
%
\begin{equation}
v=\frac{2}{1+V}.
\end{equation} 
%
%===================================% 
It is easy to see that it is a harmonic function on flat space $\tilde
\Sigma^+$, that is, 
%============<Equation>=============%
%
\begin{equation}
\partial^2 v=0.
\end{equation} 
%
%===================================% 
The boundary corresponding to the horizon is spherically symmetric. So the 
problem is reduced to the familiar one of an electrostatic potential with
spherical boundary in flat space. We 
can easily see that the level surfaces of $v$ are spherically symmetric 
in the full region of $\tilde \Sigma^+$. So we have shown that $\Sigma$
is spherically symmetric and 
then the spacetime must be the Schwarzschild-Tangherlini spacetime. 
This completes our proof. 

\section{Outlook}

We have proven a no-dipole-hair theorem for $p$-form fields with $p$ in
the range (\ref{p-condition}). The proof can be straightforwardly
extended to theories containing several electric form fields $H_{(p_i)}$ of
different rank $p_i$, each with its own coupling $\alpha_i$ to the
dilaton, as long as each of the $p_i$ satisfies (\ref{p-condition}).

As mentioned above, the upper bound on $p$ is a
natural one given the requirement of asymptotic flatness. But the
physical motivation
for the lower bound, if any, is unclear. Could static black
holes support dipoles when $p<(n+1)/2$? The answer when $p=2$ is known:
the uniqueness theorem of \cite{GIS2} affirms that a static black hole
can have electric monopole charge, but not any higher multipole.
However, here we are more interested in $p>2$ where monopoles are not
allowed. For instance, could there be static black holes in $n\geq 6$
with electric three-form, {\it i.e.}, string, dipole? The heuristic argument
presented in the introduction would seem to run counter to this
possibility, but maybe this argument misses a way to balance or cancel
the tension of dipole sources that does not involve centrifugal forces.
If this were the case it would be a striking new feature of static black
holes afforded by higher dimensions. Alternatively, and more simply,
maybe our no-dipole-hair theorem can be strengthened to rule out all
$p$-form dipoles whenever $p\leq n-1$. This issue seems worthy of
further investigation.

\begin{acknowledgments}
This work was initiated at Yukawa Institute, Kyoto University. We 
thank Misao Sasaki, Norihiro Tanahashi and Takahiro Tanaka for useful
comments. RE also thanks KIAS (Seoul) for hospitality while part of this
work was in progress. S.~O. thanks Professor Takashi Nakamura for his
continuous encouragement. This work was supported by the Grant-in-Aid
for the Global COE Program from the Ministry of Education, Culture,
Sports, Science, and Technology (MEXT) of Japan. R.~E. is partially
supported by DURSI 2009 SGR 168, MEC FPA 2007-66665-C02 and CPAN
CSD2007-00042 Consolider-Ingenio 2010. T.~S. is partially supported by
from the Ministry of Education, Science,
Sports, and Culture of Japan, Grant-in-Aid for Scientific Research Nos.~21244033,~21111006,~20540258, and
19GS0219, the Japan-U.K. Research Cooperative Programs. 
\end{acknowledgments}

%---------   References   ---------%

%---------   References   ---------%

\end{document}